# The Enron Corpus: Where the Email Bodies are Buried?

Dr. David Noever
Sr. Technical Fellow, PeopleTec, Inc.
www.peopletec.com
4901-D Corporate Drive
Huntsville, AL 35805 USA
david.noever@peopletec.com

**Abstract**
To probe the largest public-domain email database for indicators of fraud, we apply machine learning and accomplish four investigative tasks. First, we identify persons of interest (POI), using financial records and email, and report a peak accuracy of 95.7%. Secondly, we find any publicly exposed personally identifiable information (PII) and discover 50,000 previously unreported instances. Thirdly, we automatically flag legally responsive emails as scored by human experts in the California electricity blackout lawsuit, and find a peak 99% accuracy. Finally, we track three years of primary topics and sentiment across over 10,000 unique people before, during and after the onset of the corporate crisis. Where possible, we compare accuracy against execution times for 51 algorithms and report human-interpretable business rules that can scale to vast datasets.

## Introduction

The 2002 Enron fraud case uncovered financial deception in the world's largest energy trading company and at the time, triggered the largest US bankruptcy and its most massive audit failure [1]. For the previous six years, *Fortune* magazine had named Enron "America's most innovative company." By 1999, Enron was brokering a quarter of all electricity and natural gas deals [1]. Piggybacking on the internet bubble, Enron devised methods to sell everything and own nothing. Problematically, the company could assign its own revenue (mark-to-market) and then bury its losses using off-book debt in shell companies (or Special Purpose Entities called Raptor and Chewco). Summarizing the corporate state on May 10, 2000, Enron employee Susan Scott emailed, "This job is absolutely crazy. If you ever watch 'ER,' that is usually what my day looks like, only without all the blood."

| Executive Role | First Name | Last Name | Primary Email | Aliases |
|---|---|---|---|---|
| CEO | David | Delainey | david.w.delainey@enron.com | 7 |
| CEO | Kenneth | Lay | kenneth.lay@enron.com | 6 |
| CEO | Jeffrey | Skilling | jeff.skilling@enron.com | 7 |
| Energy Trader | Timothy | Belden | tim.belden@enron.com | 4 |
| Sr. VP Broadband | Scott | Yeager | scott.yeager@enron.com | 3 |
| Sr. VP Eng.Ops Broadband | Rex | Shelby | rex.shelby@enron.com | 2 |
| CEO Broadband | Kenneth | Rice | kenneth.rice@enron.com | 3 |
| CEO Broadband | Joseph | Hirko | joe.hirko@enron.com | 0 |
| COO Broadband | Kevin | Hannon | kevin.hannon@enron.com | 3 |
| CFO | Andrew | Fastow | andrew.fastow@enron.com | 2 |
| Dir Global Finance | Michael | Kopper | michael.kopper@enron.com | 0 |
| Dir Investor Relations | Mark | Koenig | mark.koenig@enron.com | 1 |
| Treasurer CFO | Raymond | Bowen | ray.bowen@enron.com | 2 |
| Vice President | Christopher | Calger | chris.calger@enron.com | 3 |
| Chief Accounting Officer | Richard | Causey | richard.causey@enron.com | 1 |
| Chief Accounting Officer | Wesley | Colwell | wes.colwell@enron.com | 0 |
| Treasurer | Ben | Gilsan | ben.gilsan@enron.com | 3 |
| CFO Global Power Pipelines | Paula | Rieker | paula.rieker@enron.com | 2 |

**Table 1 Persons of Interest (POI) in Enron Email Corpus.** *The POI list highlights the dominance of executive financial personnel from energy, broadband and power divisions, along with many email aliases.*

On December 2, 2001, Enron Corp., the fifth-largest US company, declared Chapter 11 bankruptcy. Its peak stock value per share in 2000 plummeted from $95 to $1, thus Enron evaporated $74 billion in assets and laid off 4,000 employees directly. Nearly 100,000 indirectly [2] would lose their jobs globally when the oldest and largest US accounting firm, Arthur Andersen, collapsed after

admitting its fraudulent audits and evidence destruction. However, in the year leading up to the bankruptcy, 144 senior Enron executives received $744 million in performance incentives while 24,000 retirement accounts lost $1 billion [1]. The Chief Strategies Officer J. Clifford Baxter committed suicide and the former CEO and Chairman Kenneth Lay died of a massive coronary in 2006 while awaiting sentencing on fraud and conspiracy. As recently as 2015 [1], Enron-related litigation continued. In December 2015, the Securities Exchange Commission (SEC) received a long-fought summary judgement barring former Enron CEO Jeff Skilling from serving as a board member on any publicly held companies. In total as shown in **Table 1**, over 20 executives either pled guilty or received convictions.

In 2003, the Federal Energy Regulatory Commission (FERC) released two terabytes of Enron data consisting of emails, scanned documents and audio transcripts [3]. The email traffic spans the key investigative timeline (1999-2002) and totals approximately 1.6 million emails. Since the 2003 release of Enron data, Google Scholar lists 20,200 published research articles citing the Enron Email Corpus (905 in 2018 alone). For examining real-world fraud, this corpus represents the largest public-domain email database [4].

Fast-forwarding two decades, the data burden underlying a typical 2018 investigation has mushroomed as each person now generates a digital footprint of 500 MB per day (IBM 2012). What might have been called a fingerprint in 2002 became a digital footprint in 2012 but is now regarded as a digital lake. Scaling up the 2002 Enron online production to 2018, an average Fortune 500 company with 4000 employees might produce the same two terabytes in only one day. In preparing a two-year digital audit, current fraud case might yield 1.5 petabytes – the equivalent of indexing the US Library of Congress, or a mile-high stack of CDs available for storage in 2002. In privacy terms, one also might compare capturing a corporate email server in 2002 to what modern forensics might hoist from scanning 150 corporate smartphones. Even though email traffic was tolerated as unencrypted messaging two decades ago (and today), most users regarded their communications as both private and transactional as a primary means for conducting routine personal, family, and corporate business. Enron employees routinely shared personal information, viruses, etc. Much of this damaging data still travels the internet with each new corpus download.

The present work applies modern machine learning to four key investigative tasks. First, we identify persons of interest (POI), using financial records [5] and email [6]. Secondly, we find any publicly exposed personally identifiable information (PII) such as social security numbers, credit cards and birthdates. Thirdly, we automatically flag legally responsive emails [7] using predictive coding from a seed of expertly scored examples. Finally, we track three years of primary topics and sentiment across over 10,000 employees before, during and after the onset of the corporate crisis.

To follow the trail of deceptive and suspicious communications, previous researchers [8-14] have applied sophisticated analyses of Enron's social network. The company's high-trust or collegial network distinguishes its communication pattern from fearful or angry patterns typical of rogue traders or insider threats. Enron executives relied heavily on official auditors, lawyers and bankers to certify and validate their misdeeds. Given the dire cloud surrounding its eventual downfall, the email corpus shows more positive sentiment than one might initially expect (**Figure 1**). Trust and anticipation dominate fear, disgust, sadness and anger. Thus the four investigative tasks share less with finding sub-networks of saboteurs and more with exploring a scandal in a college fraternity or crime syndicate (e.g. Racketeer Influenced and Corrupt Organizations or RICO). Newspaper accounts at the time [1] described the cozy, trusting links between Enron and its auditors. "It was like these very bright geeks at Andersen suddenly got invited to this really cool, macho frat party," said Leigh Anne Dear, a former senior audit manager at Andersen who worked in Enron tower [1].

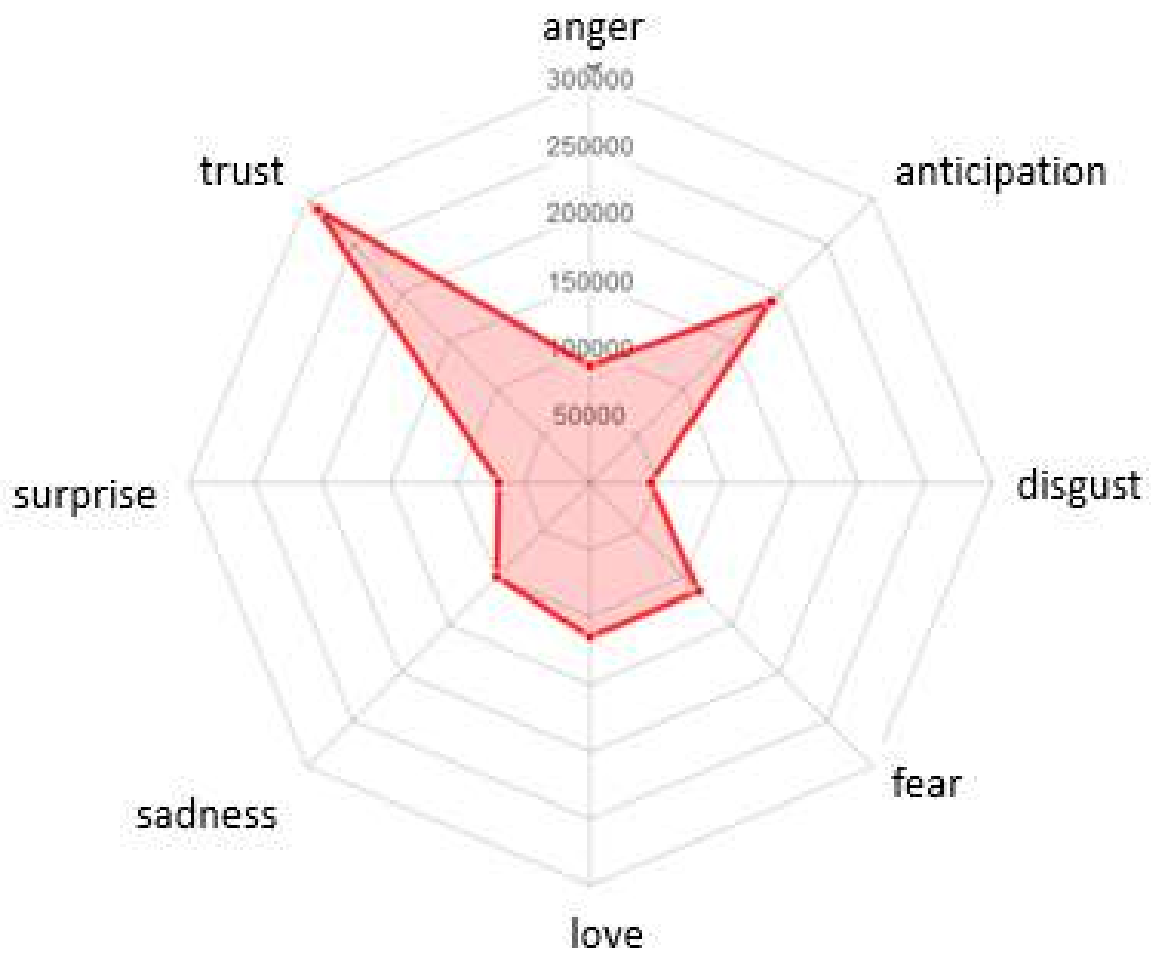

**Figure 1 Radar Chart of Emotions in Enron Email Corpus.** *Notable positive indicators of trust and anticipation with very low negative indicators for disgust, fear, anger or surprise.*

Another challenging Enron feature stems from the paucity of high-executive written communication. The key Enron figures (such as Kenneth Lay and Jeffrey Skilling) offered few email insights, either because of their corporate status generally or their hesitancy to commit their sentiment to print. Because of this loyal atmosphere and infrequent written communication between key figures (<4000 total emails), we focus less on the specific Enron social network and more on tasks for prediction and document discovery. This framework provides a general machine-learning approach to fraud detection and predictive legal coding.

**Methods**

The present work relies on three custom datasets. To render a subset called the Enron Email Corpus (EEC) previous researchers [6, 15-17] have cleansed and pre-processed the original FERC data. Secondly, as a further relevant subset of the EEC, the legal track of the 2010 Text Retrieval Conference (TREC) [7] highlighted a document discovery task to find legally responsive emails from a smaller expertly scored set of examples. For automating this discovery and classifying all responsive emails, we apply machine learning to the TREC training set. Finally, to fill in many predictive features that Enron executives' emails alone left dangling, we seek added financial insights using insider corporate compensation tables [5] such as bonuses.

***Enron Email Corpus.*** Among the many circulating Enron email datasets, this work focuses on the 2004 University of Southern California version by Shetty and Adibi [15] as subsequently updated in 2016 by Arne Hendrik Ruhe [16]. When ported to a database, their cleansed corpus includes the message content, recipient information and an employee list with some (but not all) email aliases. As shown in **Table 1**, several key figures maintained as many as seven email aliases across multiple domain accounts (e.g., msn.com, yahoo.com). We combine their four database tables into a single comma-separated value (CSV) file, removing any messages lacking either a date or sender. Because of its complex folder structure, these unsorted documents follow the timeline of single individuals but not the global corporate event sequence.

We sort by date and yield 252,607 final documents in chronological order. It is worth noting that this modified corpus represents a fraction (15%) of the actual 1.6 million emails exchanged from 2000-2002, as various legal actions have redacted and edited the original public domain version [17]. As noted by other EEC curators [18], the original dataset also reports many email fragments including machine-generated headers, embedded HTML tags, spam, automated signature appendices and many forwarded copies that lack valid initial sender information for linguistic analysis. We finally analyze this CSV file (76 million words) for the fifteen fields including standard dates, email destinations, subject and body text. We included a Boolean field for persons of interest (0 = negative or 1 = positive). The "Sender Name" field derives uniquely from all email aliases to track a single person across multiple formatted addresses and domains. The "Subject" and "Body" fields for each document represent the first-person content pertinent to sentiment analysis, since this content originated from the sender without forwarding.

***Privacy Filters.*** An important task spearheaded by Duke Law School's Electronic Discovery Reference Model (edrm.net) involves finding personal, health, and financial information (PII) [19]. In 2003 when the FERC first dumped raw Enron server logs into the public domain, the automated identification of personal data (including social security and credit card numbers) may have appeared unrealistic. However, the need to filter out such PII has only risen in urgency with the rise of large data exposure, particularly in social media [20]. It is worth recalling that investigators brought no legal actions against 99.83% of the unique enron.com email users, yet many still have exposed PII. Prior to social media, email correspondence dominated most online transactions (such as personal purchase receipts and Human Resources spreadsheets). However, even an email harmlessly congratulating a colleague on their birthday can expose enough information to generate some credit card fraud in their name. **Table 2** shows an approximate rank-order of potentially damaging personal information embedded in the EEC. Some notable challenges derive from the lack of a uniform bank account notation and state differences in driver's license (DL) formats. We opt to detect International Bank Accounts that follow the uniform IBAN format and rely on a compiled database of state DL formats. Not shown are the most common PII elements such as email or Internet Protocol (IP) addresses, which predictably would include embedded identifiers given their origins in server logs. Also not shown are matters of potential personal harm such as romantic affairs, attached photos, family disputes, and funerals or births, even though we discovered many such exposed examples.

| Damaging PII |
| --- |
| Social Security Number |
| Credit Card Number |
| Passwords |
| Passport Number |
| Driver's License Number |
| Intl Bank Account |
| Birth Date |
| Phone Numbers |

**Table 2 Rank order of damaging PII in Email**

***Legally Responsive e-Discovery and Predictive Coding.*** The legal track of the 2010 Text Retrieval Conference (TREC) [7] initially described the second important dataset. Unlike the broader accounting fraud case, Enron also faced sanctions for its involvement in the California electricity blackouts from 2000-2001. Reinforcing Enron's role, then CEO Kenneth Lay ridiculed regulatory efforts to the Chairman of the California Power Authority in 2000 [21], *"In the final analysis, it doesn't matter what you crazy people in California do, because I got smart guys who can always figure out how to make money."* The FERC eventually investigated Enron for artificially reducing grid electricity to profit when prices spiked as much as 800% [21]. Effectively, in this case, Enron bought power for $250 in California and sold it immediately to Nevada or Washington for $1,200. The FERC fined Enron $1.52 billion for their scheme, but the economic fallout of rolling blackouts has been estimated to have cost the US $40-50 billion and displaced then California Governor Gray Davis [21].

This subset of Enron emails represents 855 handpicked messages, of which legal experts scored 139 as "responsive" to the California energy bid manipulation case and the remaining 716 scored as "unresponsive". Applying machine learning automates the army of legal clerks reading and scoring emails to include in the fraud file. As also called "technology-assisted reviews (TAR)" or computer categorization, this task provides learning by example and automates the kind of triage tasks done typically to prepare document discovery prior to legal filing. We bootstrap the derived classification rules to score the entire quarter-million emails and identify possible previously missed but responsive emails.

From the original corpus of hand-scored emails, we initiate a pipeline of text analysis by first transforming the input. We lowercase the email text, remove punctuation and common English terms ("stop words" such as "the" or "a"), strip white space, then stem each word (e.g. transform "bidding" to "bid" using the extended Porter algorithm or Snowball stemmer). We convert this transformed corpus into a document term matrix where the rows are documents and the columns are weighted word frequencies. The weighting transform provides the "term frequency – inverse document frequency" (TF-IDF), a numerical statistic that increases the more times a word is used in a single email offset by its frequency in other emails. As a measure of an email's uniqueness, an example high TF-IDF score for "California" corresponds with many mentions compared to few mentions in other emails. This corpus of 85 documents yields 21877 unique terms. We further simplify the resulting sparse matrix to remove terms that do not appear in at least 97% of the emails. The resulting final matrix consists of 789 differentiating terms. We randomly sample and split 30% these emails (or 257 rows) into an out-of-sample test set to verify the training on the other 70% cases.

This e-Discovery example corresponds to a highly unbalanced dataset (5:1). To rebalance the responsive and unresponsive classes, we create two variants to correct the fivefold preponderance of unresponsive messages. We train on both these over-sampled and under-sampled datasets. The resulting under-sampled training data consists of 97 examples of both cases (responsive and unresponsive), while the over-sampled training data consists of 501 examples of both cases. We train six families of algorithms to distinguish responsive from non-responsive emails: 1) neural network; 2) random forest; 3) k-nearest neighbors; 4) recursive partitioning and regression trees (CART) ; 5) gradient boosting and 6) boosted classification trees. **Figure 2** shows the resulting performance optimization. We compare both the accuracy and relative execution time to train with three-fold cross-validation on both the under- and over-sampled data. All algorithms achieve 80% or greater accuracy. The fastest and most easily explained algorithm is a traditional decision tree (or CART) and we will deploy this for high-throughput legal document discovery in the Results section.

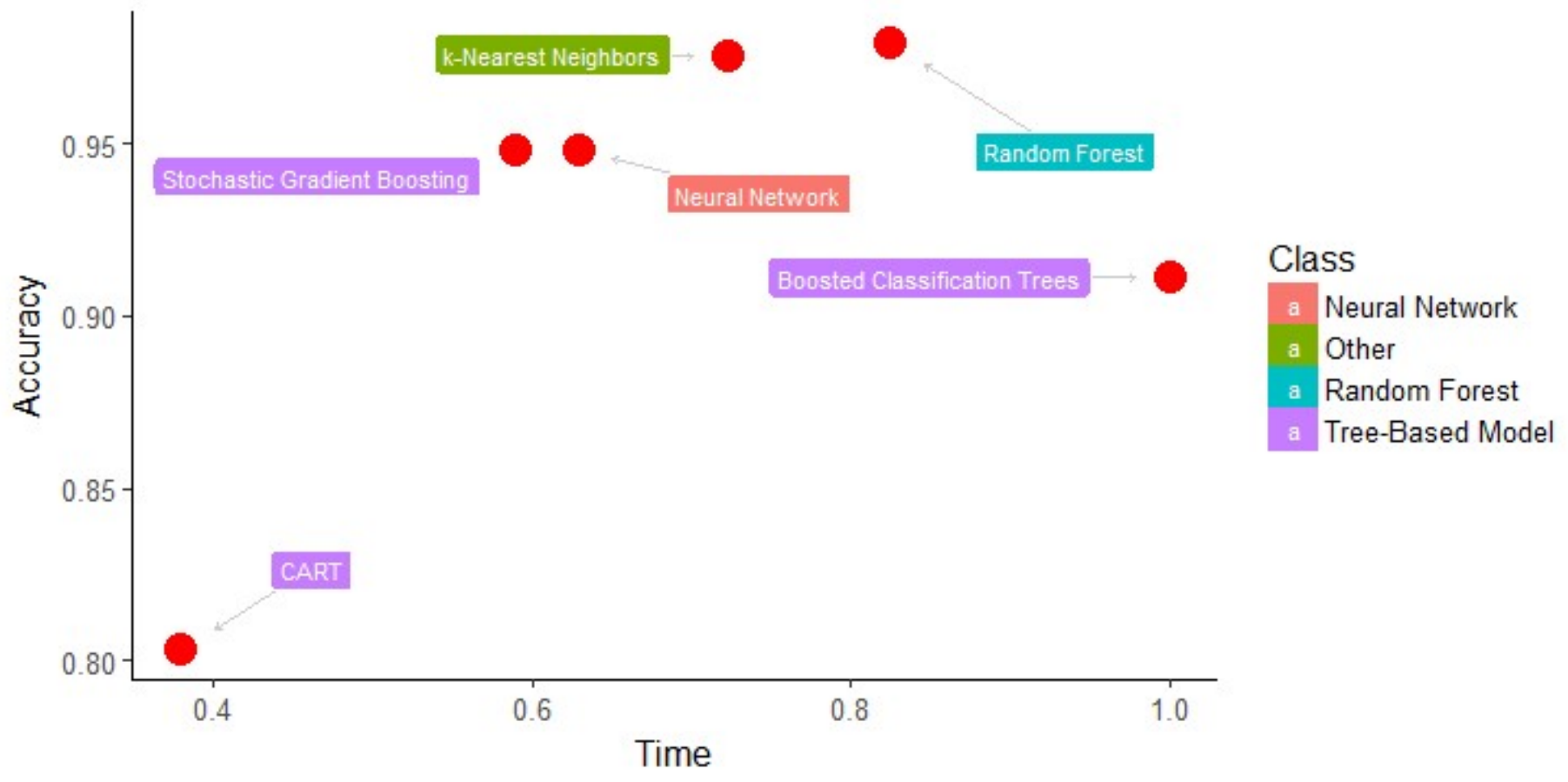

**Figure 2 Performance and Accuracy for Classifying Responsive Emails.** *The most accurate algorithm is Random Forest; the fastest one is CART that is traditionally the easiest to interpret in explainable legal rules.*

***Insider Compensation.*** As part of insider financial disclosures [5], a final training dataset details the salary, bonus, and both the short-term and long-term compensation structure for 145 Enron executives. The dataset includes over 20 possible financial predictors of fraud, but offers a challenging sparse matrix to fill in any blanks or not-applicable values. For instance, out of 3358 possible values, over 40% of cells are blank (1356/3358). For machine learning purposes, we opt to fill these blanks with zero values, although alternatives (such as replacing all blanks as medians or averages) do not alter the main decision tree or the relative importance of each feature to concluding an Enron employee likely was a Person of Interest (POI).

***Sentiment and Personality Trait Linguistics.*** Since many lone-wolf insiders present the opposite traits seen in the Enron fraud case, we apply emotional and linguistics tools to quantify Enron's unique and tight-knit nexus of internal-trust. The two most popular terms in the email corpus are predictably "Enron" and "our", reinforcing Enron's shared, collective endeavor. This loyal network of executives sent few and terse emails. We analyze sentiment factors in email in a multistep process. We account for single word stems by removing numbers, stripping punctuation and lowercasing the text. For English sentiment, we employ the AFINN-111 [22] scale for positive or negative polarity. Examples of strong negative sentiment include most curse words, racial or class epithets and grievances (e.g. "tortures"). Examples of strong positive sentiment include broad approvals (e.g. "breakthrough" or "outstanding"). We also apply alternative crowd-sourced dictionaries such as Bing and the Word-Emotion Association Lexicon with eight basic emotions (anger, fear, anticipation, trust, surprise, sadness, joy and disgust).

***Distributed Processing Framework.*** To enable distributed and parallel processing of larger datasets (terabyte or larger) across computing clusters, we transferred the Enron emails to the Apache Hadoop [23] software suite (Cloudera® CDH Virtual Machine 5.13.0). We deploy multiple detection algorithms using Hadoop streaming, which enables rapid code modifications and agile developer customizations specific to each training case. We package each cluster as a virtual machine image (Oracle® VirtualBox) [24]. The open source framework of this underlying pipeline allows resource-sharing independent of hardware, guest operating systems and complex version requirements. In 2018 [23], some of the larger clusters (e.g. Facebook, Yahoo) may total 40,000 Hadoop nodes to handle petabytes of distributed file storage.

***Machine Learning Framework.*** To enable predictions based on Enron email narratives, we apply an ensemble of machine learning methods, including recursive partitioning (simple decision trees), support vector machines (SVM), neural networks, boosted and bagged tree models such as random or isolation forests. To account for the unbalanced datasets (where negative classes outweigh positive ones), we resample and report not only accuracies on training, validation and test sets but also employ metrics that are insensitive to the lack of balance, such as receiver-operator characteristic curves (ROC, or true vs. false positives), sensitivity (true positive rate) and specificity (true negative rate). In the absence of rebalancing, most algorithms converge easily to 100% benign activities, given few overall malicious acts. We export the results as JavaScript Object Notation (JSON) files for post-processing and downstream visualization.

**Results**

We report our Enron data mining and machine learning tasks principally on metrics of accuracy, execution times and human understandability criteria. We extract where possible example new findings for the four major tasks outlined in the Methods section.

***Entity Extraction of Personal Information.*** From their initial Enron analysis, the Duke Law School's EDRM [19] account of PII reported 10,000 separate information elements, including 60 credit card numbers, 572 social security Numbers (SSN), and 292 individuals' dates of birth. The present work identifies nearly 50,000 additional items that might qualify as PII in **Table 3**, including potentially 86 new credit cards, 30 passport numbers, 18 international bank accounts and 4 social security numbers. **Table 3** summarizes these results for extracted PII and shows that 96% of these new findings represent potentially interesting dates and phone numbers.

| PII Description | Count of Item | Percent |
|---|---|---|
| Date | 24283 | 49.47% |
| Phone | 23002 | 46.86% |
| Drivers License | 1663 | 3.39% |
| Credit Card | 86 | 0.18% |
| Passport | 30 | 0.06% |
| Intl Bank Account | 18 | 0.04% |
| SSN | 4 | 0.01% |
| Grand Total | 49086 | 100.00% |

**Table 3 Count of Potential Enron Employee PII Embedded in Corpus**

The 86 credit card related emails include all major issuers (AMEX, VISA, MasterCard) and multiple sources of PII (expirations, names and mailing addresses). As circulating in both the original 2003 and cleansed 2016 Enron email corpus, several clear examples of raw credit card information carry severe privacy consequences, such as "*it is a Mastercard and the number is XXXX XXXX XXXX XXX and the exp is 11/01*". Other examples provide available credit limits, such as one email from a senior executive describing his impending Atlantic cruise: "*What kind of a fool do you think I am. No way I am getting in the middle of you and Amanda on vacation plans. ..I will however give you both a credit card number - it is XXXX XXXX XXXX XXX… You and Amanda have a Caronia class cabin at $2045.32 per person*." One eBay auction email confirmed the winning bid for Dave Mathews tickets and supplies VISA card number, expiration and mailing address. Another asked for two University of Texas season football tickets with VISA information, *"I thought given the short time frame I would just email with my information...My credit card information is XXXXXX, Visa. Thanks for your help."*

One particularly troubling travel agency email was copied multiple times to a hotmail account and begins with subject "*Re: Paris*" and the body text reads, "*Here are the list of roommates for each room with one credit card… Unless otherwise specified, all credit cards are in the name as listed below..*", and then proceeds to list names, card types, card numbers, and expirations for eight individuals, including several cards unexpired at the time of public EEC release (3/26/2003).

These new findings contradict the official FERC timeline for Feb. 13, 2002, which stated, "Eventually, FERC collected over 2.2 terabytes of information, which it made available to the public, *with the exception of the small amount of data that contained social security numbers and personal information not related to the investigation* [25]." Furthermore, even with expired cards, many companies even today will reissue the same numbers many times with new increments for expiration and Card Verification Value (CCV). Even when not charged, the mere existence of credit and debit cards in public circulation poses risk for account hijack, account recovery protocols (e.g. Amazon), and conceivably identity theft.

From this time (1999-2001), one common, but risky email practice would embed passwords for various accounts directly in a link (e.g. http://somedomain/myaccount?userid=NNNNN&password=XXXXX). Our examination reveals 70 such examples, mostly from benign newsletters or Enron's United Way campaigns. Often these passwords arose within bulk email services (like cheetahmail.com) for retailers (neimanmarcus.com). These third parties additionally provided unencrypted link requests to Enron employees as http, not https, a practice commonly exposing such passwords to plaintext eavesdropping. As the worst outcome, these sloppy practices led to unwanted spam from guest passwords. One set of employees did however receive software source code to an Oracle database connector with plain text login (user=user, password=password), a case where these credentials likely reflected harmless anonymizing but which unfortunately also show up in actual careless applications.

For many of these PII categories, one expects high false positives, particularly given either the formatting ambiguities or the overlap with other common embedded numbers. Examples include nine-digit zip codes that run together to mimic social security numbers. However, even this particular false positive carries some irony, as recent research [26] has demonstrated that identity thieves can exploit such zip codes alone to predict narrow ranges of individual SSNs when correlated with other public records. For example, for individuals born after 1988, the first five digits can be matched 44% of the time using only zip code and public birth dates [26]. Both for this PII discovery task and for predicting legal relevance, each selected algorithm should carry a greater weight to short-list potential positives, particularly given the higher costs associated with missing an actual hit compared to over-reporting.

***Predicting Legal Relevance.*** In the TREC 2010 document retrieval competition [7], twenty teams submitted predictions with an average accuracy of 59.5% and a range from 18-99%. As shown in **Figure 2**, our best algorithm (random forest) yielded 99% accuracy.

| Metric | Description | Value |
|---|---|---|
| False Positive Rate | Mislabeled Responsive | 9.30% |
| False Negative Rate | Mislabeled Non-responsive | 40.50% |
| Sensitivity | True positive rate | 59.50% |
| Specificity | True negative rate | 90.70% |
| Accuracy | TP + TN/Total | 85.60% |
| Error Matrix | Predicted | |
| Actual | Responsive | Non-responsive |
| Responsive | 25 (TP) | 17 (FP) |
| Non-responsive | 20 (FN) | 195 (TN) |

**Figure 3 Decision Tree Metrics for Predictive Legal Coding.** *Overall accuracy of 86%. The yellow highlighted boxes correspond to raw email counts accurately classified.*

As a follow-on, we opted to explore a simpler decision tree model because of its human interpretability as concise business rules. As shown in **Figure 2** and **3** together, it offered a two-fold speed advantage but only slightly less accuracy (86%, **Figure 3**). Remarkably, using only six keywords (California, demand, bid, system, gas and Jeff) among 76 million total EEC words, we can equal the legal scores from human experts. As shown graphically in Supplemental Material, **Figure 7** and textually in **Figure 4** the resulting seven business rules provide an efficient automation path, particularly for parallel processing of huge datasets without an army of legal interns.

```
Non-responsive classification
Rule 1: when california <  1.5 AND demand <  0.5 AND bid <  0.5 AND gas <  1.5 then non-responsive (0)
Rule 2: when california <  1.5 AND demand <  0.5 AND bid <  0.5 AND gas >= 1.5 AND jeff <  0.5 then non-responsive (0)
Rule 3: when california <  1.5 AND demand >= 0.5 AND system <  0.5 then non-responsive (0)

Responsive classification
Rule 4: when california >= 1.5 then responsive (1)
Rule 5: when california <  1.5 AND demand <  0.5 AND bid <  0.5 AND gas >= 1.5 AND jeff >= 0.5 then responsive (1)
Rule 6: when california <  1.5 AND demand >= 0.5 AND system >= 0.5 then responsive (1)
Rule 7: when california <  1.5 AND demand <  0.5 AND bid >= 0.5 then responsive (1)
```

**Figure 4 Classification Business Rules for Legal Responsive Enron Emails**. *Seven rules can classify 86% of the responsive email discovery using six keywords.*

Applying just the positive or responsive rules (4) to the entire quarter-million emails, we identify 13,248 (5%) new emails as relevant to the legal case for California market bids [21]. Although the TREC competition involved a slightly different EEC corpus, the average yield of responsive documents (8%) compares favorably to our simple business rules summarized in **Figure 4**. One noteworthy observation stems from the word importance of the name “Jeff” as a defining rule in legally responsive emails. While Enron CEO Jeff Skilling denied California price manipulation, he pushed employees to "trade aggressively", and if unwilling then to "find another job".

One additional example of an email automatically identified outside of the initial TREC sample relates to the California price manipulation scheme directly. An outside legal consultant sent an email to 15 Enron employees on January 29, 2001, offering effectively a call for written self-incrimination: "*Steve thinks he might be asked about whether the market was manipulated. Please provide information on whether this was the case and who the participants likely were.*"

***Predicting Persons of Interest.***
*Financial Factors.* One significant predictor of a given employee's likelihood of conviction was their compensation structure. We tested the 145 employees included in public financial disclosures [5] and tested whether financial compensation (both short- and long-term) could correctly distinguish between eventual persons-of-interest or not. Since the financial dataset is smaller than the email corpus, we applied an extended suite of 51 machine-learning algorithms in 16 major families (Bayesian, linear, polynomial, neural nets, support vector machines, and tree-based models). We scored each algorithm's accuracy and relative execution time (**Figure 9** in the Supplemental Material). For inaccurate, lengthy training, the worst algorithm for these smaller datasets turns out to be the currently popular stacked auto-encoder deep neural network, which tends to require millions of training cases.

The performance plot in **Figure 9** shows that a random forest algorithm achieved the highest (95.28%) accuracy of correctly identified persons of interest in the over-sampled dataset and 72.2% in the under-sampled case. The random forest also gave the rank-order variable importance. The most predictive factor of criminality proved to be whether the employee exercised stock options, followed in importance by the total Enron stock value held in their name (see Supplemental Material, **Figure 10**). This algorithmic result matches with reports of widespread cashing out by senior executives (particularly Lay, Skilling, and Fastow) during the 2001 autumn that preceded Enron's bankruptcy filing [1]. Skilling initiated sale of $65 million worth of his shares on September 6, 2001, which eventually he executed on September 17. He later claimed the terrorist attacks on 9/11 prompted his sale [28].

As seen in the previous predictive legal coding task, the decision tree (CART) and k-nearest neighbors supplied the two fastest algorithms. The simplest single branch decision tree identifies the persons of interest with 87% accuracy if they received a bonus of more than $1.17 million (**Figure 5**). In the first four months following the Enron bankruptcy, prosecutors paid particular attention to these one-time bonuses [29] as powerful incentives to manipulate revenues and inflate Enron stock. Like many other financial instruments used by Enron, these bonuses (called Performance Unit Plan (PUP)) proved complex in practice. Because the employees knew almost four years in advance they might be selected for this one-time payout and because they also knew that to make the final list implied that they generated revenue, employees who got bonuses necessarily joined in the now-discredited mark-to-market accounting. In effect, getting the PUP bonus meant an executive had to collude in Enron's accounting high-jinx. The intriguing identification of this single factor as predictive by machine learning echoes many of the same prosecutorial observations offered at the time since such a "strong financial motive is probably the best evidence a prosecutor can get to promote or to establish criminal intent" [29].

| POI | Predicted | |
|---|---|---|
| Actual | FALSE | TRUE |
| FALSE | 116 | 11 |
| TRUE | 8 | 10 |

**Figure 5 Error Matrix for POI from Financials**

*Email and Financial Factors*. We encoded several factors in the previous financial dataset that counted miscellaneous email features, such as the total to and from emails as well as the number of messages to or from a person of interest. These factors proved to be of moderate to low predictive value, as shown using the rank order method from the random forest algorithm (see Supplemental Material, **Figure 10**). One might anticipate this low predictive value given the low number of executive emails generally when coupled with the high number of executives eventually deemed persons of interest. This lowered importance becomes more obvious given that the simple decision tree achieves 87% accuracy without including any email factors. Previous work, however, has pointed to the potential to incorporate detailed linguistic analysis to predict fraudulent corporate filings (10-K) [30]. The authors also rank-order factors that companies use when deceiving investors, including the sentence length, uncertainty markers and proper nouns. We supplement the financials with email sentiment indicators, such as scored expressions of trust, deceit, fear and sadness. As shown in Supplemental Materials, **Figure 12**, the most important non-financial Enron indicators for persons of interest were total email count, fear, deception and surprise. For this case, we do not try to balance the positive and negative instances in the training data but instead choose a metric that compensates for the fewer number of bad actors. One such measure is AUROC, or the area under the receiver operating characteristic curve. For a binary classifier, this means the probability that our classifier will rank a randomly chosen positive (person-of-interest) higher than a randomly chosen negative (innocent). The best random forest model achieves 0.8625 for AUROC on the testing data. The best ruleset for a single decision tree follows in **Figure 6**.

```
Person of Interest Classification
Rule 1: when bonus >= $550,000 AND from_messages < 520 AND anger >= 10.5 then person of interest (1)

Not Person of Interest Classification
Rule 2: when bonus >= $550,000 AND from_messages < 520 AND anger < 10.5 then not person of interest (0)
Rule 3: when bonus < $550,000 then not person of interest (0)
Rule 4: when bonus >= $550,000 AND from_messages >= 520 then not person of interest (0)
```

**Figure 6 Classification Business Rules for Person of Interest from Financials and Enron Emails.** *Four rules can classify persons of interest with AUROC = 0.52 with anger as the dominant sentiment factor.*

*Email Factors*. As a final variant on predicting persons of interest, we remove all financial inputs and train on the email factors alone. One reason to seek this solution stems from the lack of real-time financial data in the early stages of a fraud investigation, particularly for non-public companies. One would like the ability to get ahead of such deceptive events, or at least monitor developing fraud as it happens without resorting to forensic tools alone. Thus, we repeat the basic machine-learning framework in the previous sub-tasks but remove all financial factors such as bonuses and salary information. We predict persons-of-interest from linguistic factors alone. The best algorithm (random forest) loses some potential accuracy, as measured with AURUC = 0.76. Stated plainly, this approach stands a 76% chance of distinguishing a person-of-interest from a random pool of mixed positive and negative cases. **Figure 14** in the Supplemental Materials shows that other than the number of emails, the most predictive emotional content in order are sadness, deception, fear, and surprise. Generally, the least important linguistic factors are the raw positive and negative sentiment (as measured by AFINN or Bing dictionaries of valence). **Figure 7** summarizes the simplest business rules for detecting persons of interest from email factors alone in a single decision tree method.

```
Person of Interest Classification
Rule 1: when fear >= 15 AND from_messages < 697 AND from_poi_to_this_person>=28.5 then person of interest (1)

Not Person of Interest Classification
Rule 2: when fear >= 15 AND from_messages < 697 AND from_poi_to_this_person < 28.5 then not person of interest (0)
Rule 3: when fear >= 15 then not person of interest (0)
Rule 4: when fear >= 15 AND from_messages >= 697 then not person of interest (0)
```

**Figure 7 Classification Business Rules for Person of Interest from Enron Emails.** *Four rules can classify persons of interest with AUROC = 0.56 with fear as the dominant sentiment factor.*

***Predicting Crisis Timelines from Email.*** In **Figure 8**, the analysis of corporate-wide email sentiment over time highlights the overall positive sentiment (above the red line), but shows a steep decline or dip as the crisis begins to dominate Enron's timeline. A more robust individual view specific to the persons of interest shows what dominated their internal conversations in **Figure 9**. In the format of a word cloud, color, size and circular segments show each individual's most frequent word choices. Each labelled segment of approximately 20 degrees in **Figure 9** shows the employee name and most frequently used email terms. The label for "Employee Benign" represents a random subset of all employees not included in the persons of interest list.

These word frequencies graphically highlight what ultimately became the focus of charges against each POI. For instance, Dan Boyle (upper right, **Figure 9**) frequently discussed Credit Suisse First Boston ("CSFB") and that connection to Special Purpose Entities (SPE) became central to the charging documents [31] for conspiracy and wire fraud. The analysis of Enron Treasurer, Ben Glisan, shows similar frequent associations with the SPE called Raptor. CEO Kenneth Lay attempted a last ditch sale of Enron to energy rival, Dynegy, and his email traffic reflected that link. As Enron founder, Lay also proved the only employee who wrote frequently of “our” and “company”.

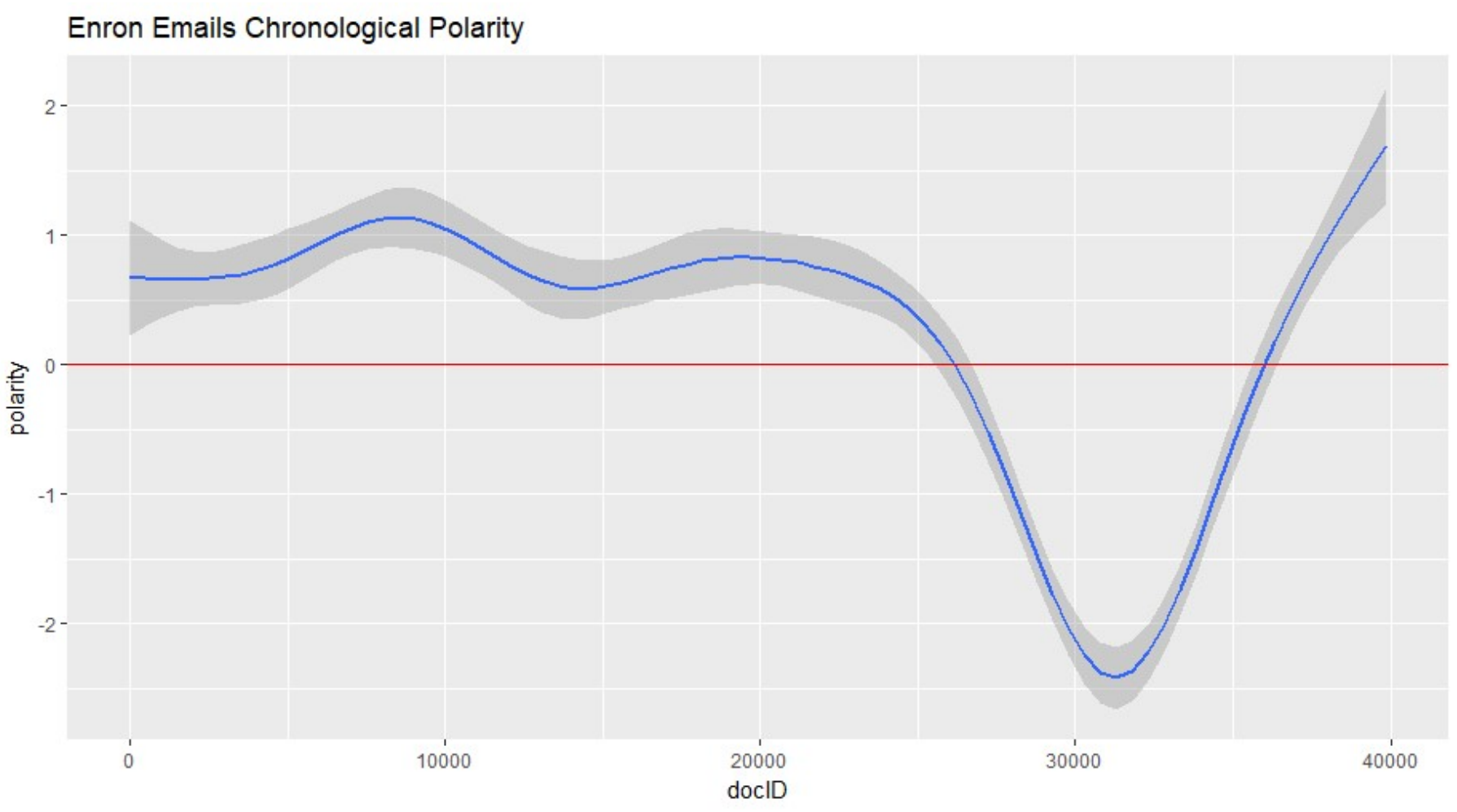

**Figure 8.**

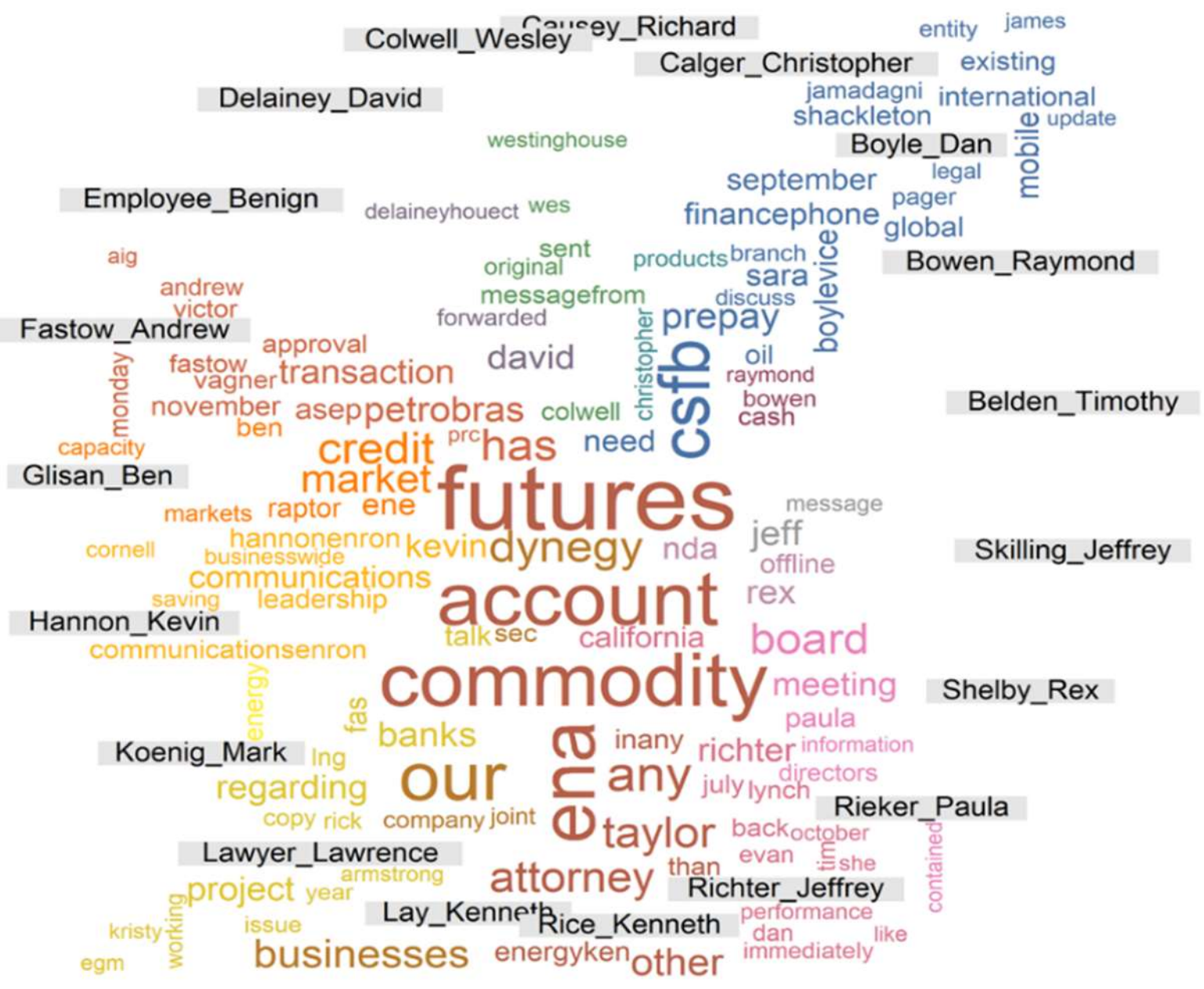

**Figure 9. Word cloud representation of Enron POI.** *Notable word frequencies match to eventual charging documents.*

A further clustering of similar emotional tenors in email can define which persons of interest share the same basic styles. As illustrated in Supplemental Material, **Figure 16**, employees share stylistic similarities that can provide additional insights beyond just social network links. For example, the email corpus from Ken Lay stands out as unique among other persons of interest. Lay effectively communicates as an outlier compared to all his subordinates. Tim Belden, managing director of Enron Energy Trading, writes the most like a generic benign employee. The persons of interest that share the most linguistically with CEO Jeff Skilling include the investor relations team (Paula Rieker and Mark Koenig), which hints at their shared and publicly facing focus on Enron stock price. The bulk of Enron’s financial engineering (right half, **Figure 16**) includes a connected cluster of treasurer, chief financial officer, and investor relations.

As shown in **Figure 9**, CEO Jeff Skilling and CFO Andrew Fastow both appear as somewhat rare Enron POIs who signed their emails most frequently. Ironically, one post-Enron reform (Section 302, SOX, or Sarbanes-Oxley Act of 2002) included a biting provision that both CEO and CFO certify in writing their SEC financial filings. SOX arose primarily to counteract one of Skilling’s defenses that he knew nothing of Enron’s off-book debts.

**Discussion and Conclusions**

Future work should consider not just the email and financial data but supplement these conclusions with the varied email attachments. Those Enron files include 43,401 files include the range of 32,000 MS Office and 9,589 Adobe PDF files (7.9 GB) [32] and serve as added enterprise inputs typical of an intranet or file server. At least one example of detectable malware is still included in the Enron corpus (called “*Joke-StressRelief*”) along with 231 executables.

A further investigative task would including automating timeline and more extensive topic construction, particularly to develop new hypotheses such as “who knew what and when?” Temporal predictions differ from the broad classifications presented here and include not only malicious actions but also actors whose behavior changes subtly over time. Critical to establishing such temporal patterns however is setting the characteristic period (or lag) and automating feature extractions when calculating baseline and rolling statistics. The simplest time-dependent fraud model might tabulate weekly statistics of user behavior with z-score statistics, or current values minus mean divided by standard deviation. This metric over time highlights outliers and changes in an individual’s behavioral pattern.

One noteworthy caveat to the present work focuses on the neglected email traffic (85%). What metrics might differ in a wider email scope including attachments and redacted data? Because machine learning ultimately relies on many statistical features, one empirical test might look for whether the EEC subset introduces unexpected statistical changes. For example, does this subset follow an expected distribution like Benford’s first-digit law that many real-life datasets over-represent small numbers? For example, a random process generates a distribution for each of the numbers (1-9) with 11.1% probability. However many natural processes heavily favor small numbers (1-3), with the number one represented with 30% probability, or three times its expected random selection. Also called power laws, these validity checks can detect accounting fraud if forensic audits fail to find numerical patterns that abides Benford’s law [34]. We ran a quick check as shown in **Figure 17** (supplemental material) against the Enron subset and verified that both its email body length and daily email counts follow the expected Benford distribution. This suggests that ignoring redacted emails generally does not alter the statistical expectations of a larger corpus. Finally because many machine learning pipelines naturally subset- either for out-of-sample tests or as built-in efforts to combat over-training—some fractional datasets actually improve the overall performance when generalizing rules to their much larger parent corpus.

As presented here, we find that the largest public domain email repository offers fresh insights when deployed in modern computing clusters and comprehensive machine learning algorithms. Despite previous claims of removing PII by both the FERC and Duke Law School, we discover missed and potentially damaging examples of innocent but exposed Enron employees. Our approach also highlights the potential to automate mundane and error-prone legal discovery tasks. This predictive coding generalizes the underlying rules discovered from a much smaller subset of expertly scored emails marked as legally responsive. The best algorithms meet or exceed the competitive results from twenty other commercial and academic data science teams. According to the American Bar Association (2012), for every 340,000 documents preserved for litigation, one actually gets used yet its discovery comprises half the total litigation cost [35]. For international cases, language translation costs further inflate this number but various rough orders of magnitude have benchmarked the market impact of US legal discovery between $30-50 billion [36]. If this task was ranked based on its own economic value, it would outrank 79 out of 189 countries.

With 95.7% overall accuracy, we finally applied our comprehensive algorithm suite to identify the main persons-of-interest by using features from both financial compensation tables and deep sentiment in email. Notable outcomes include simple business rules particular to the Enron case that could easily scale to much larger datasets with billions of documents. While trained algorithms offer novel and previously unpublished approaches, one key advantage here is the rank ordering of each factor's importance. Given the complexity of the overall task and also the need to explain or motivate the algorithm's decision in a legal framework, such knowledge of which rules and key factors contribute to make a final determination often outweighs the accuracy of prediction success alone.

**Acknowledgements**
The authors would like to thank the PeopleTec Technical Fellows program for encouragement and project assistance.

## Supplemental Material

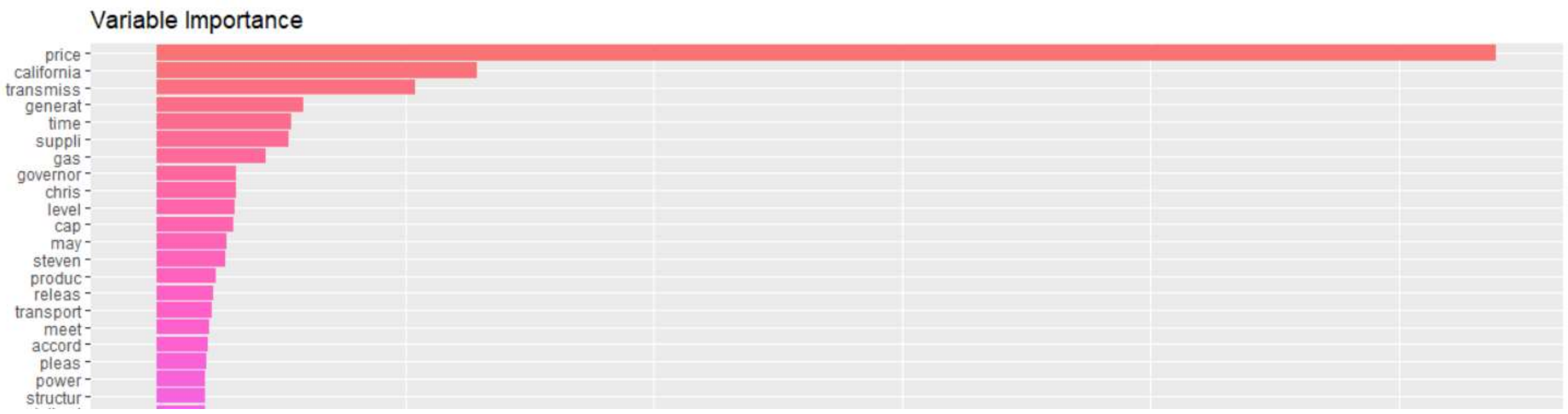

**Figure 8 Predictive Importance of Each Term for Determining Legally Relevance or Responsive Email**

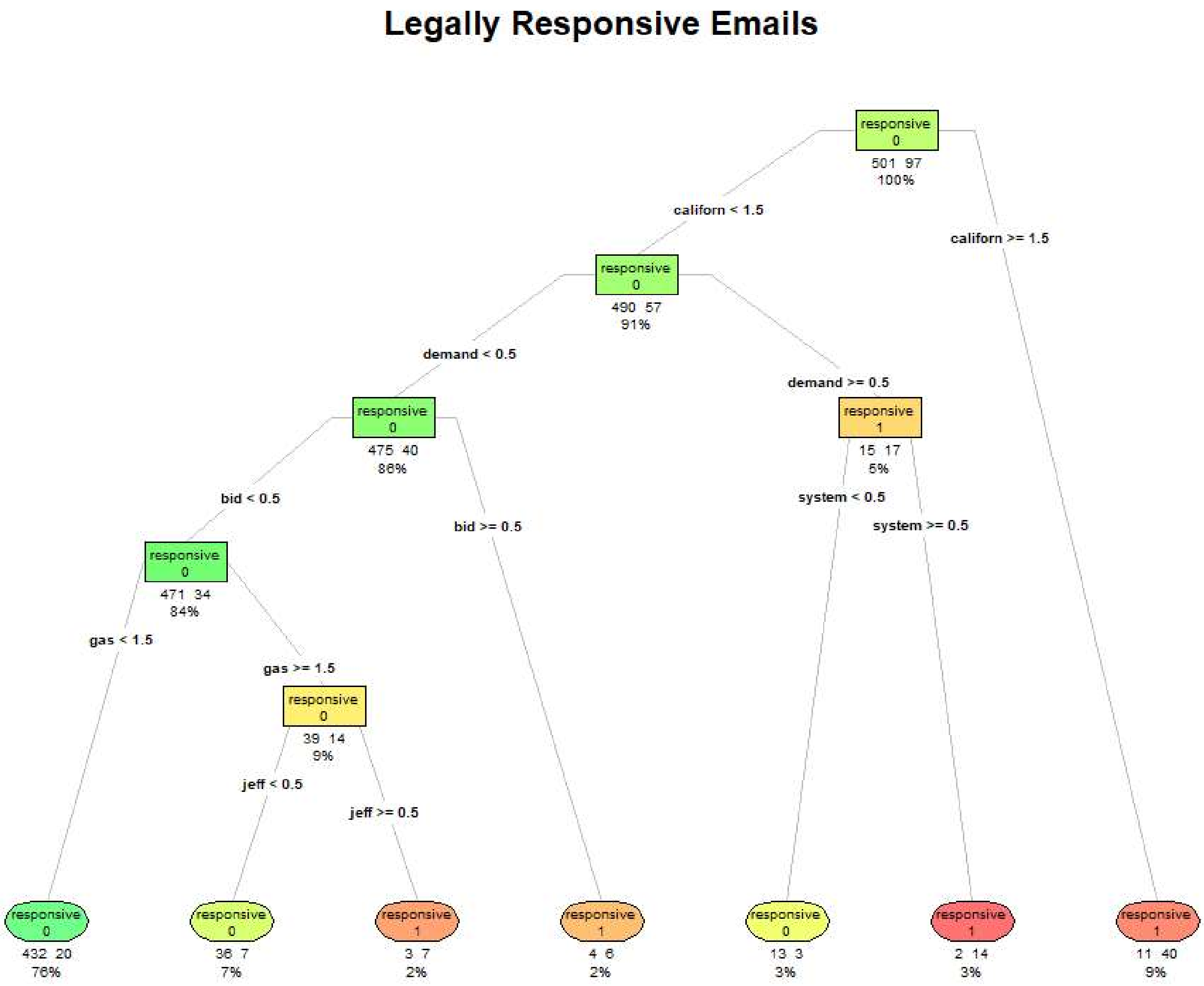

**Figure 9 Decision Tree Determining Legally Responsive Emails**

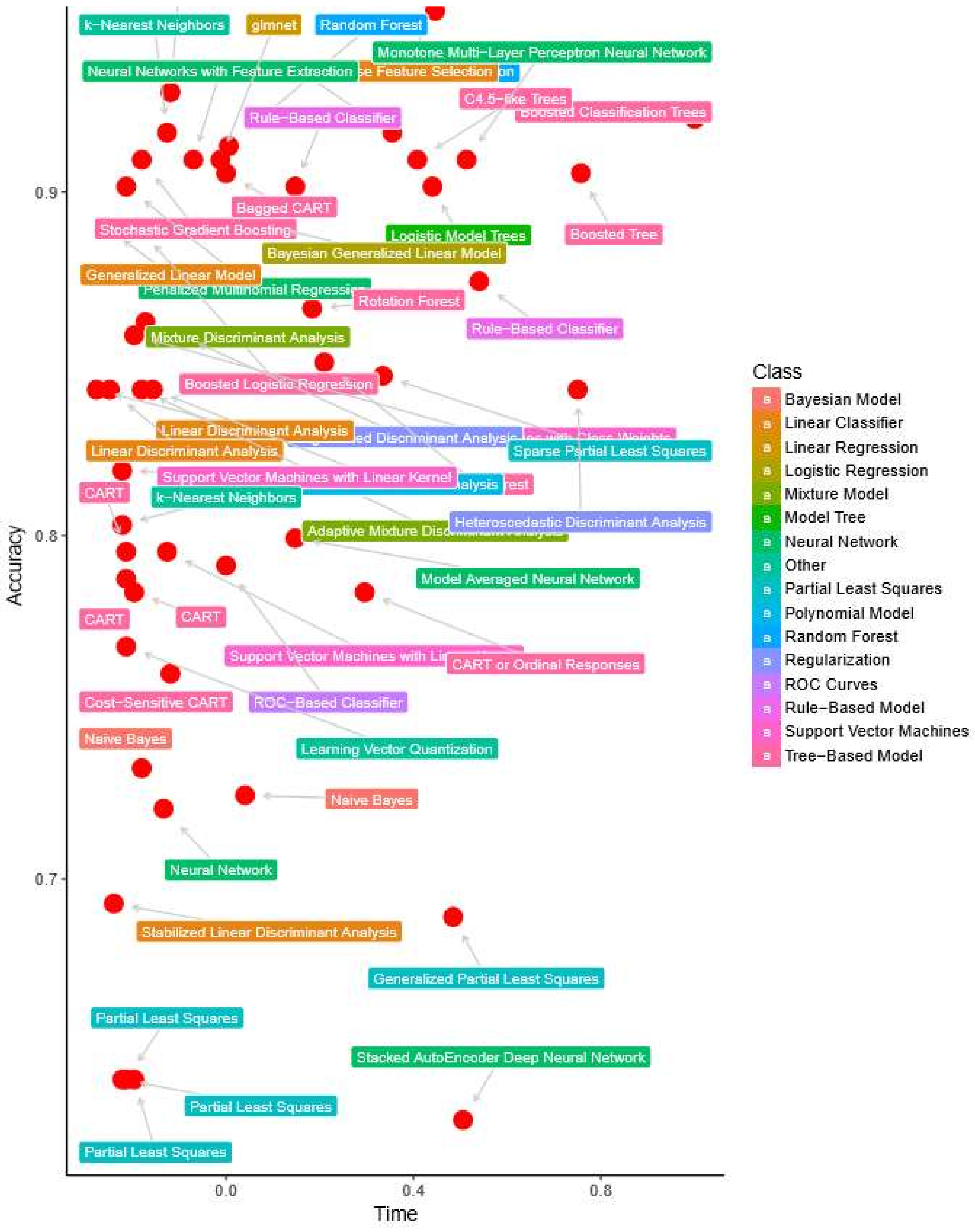

**Figure 10 Algorithm Accuracy vs. Time for Predicting Persons of Interest from Financial Compensation**

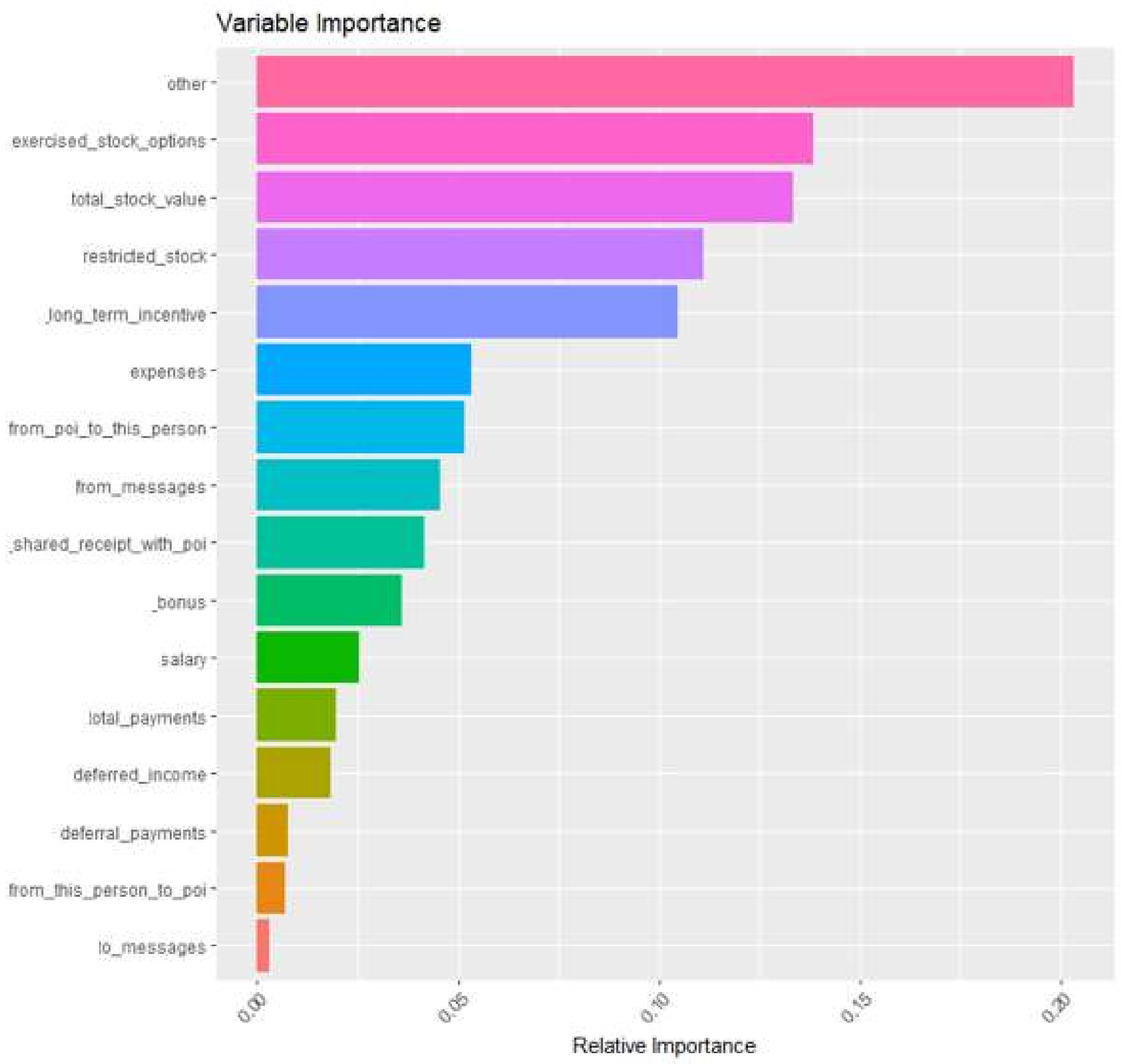

**Figure 11 Variable Importance for Predicting Persons of Interest from Financials Alone**

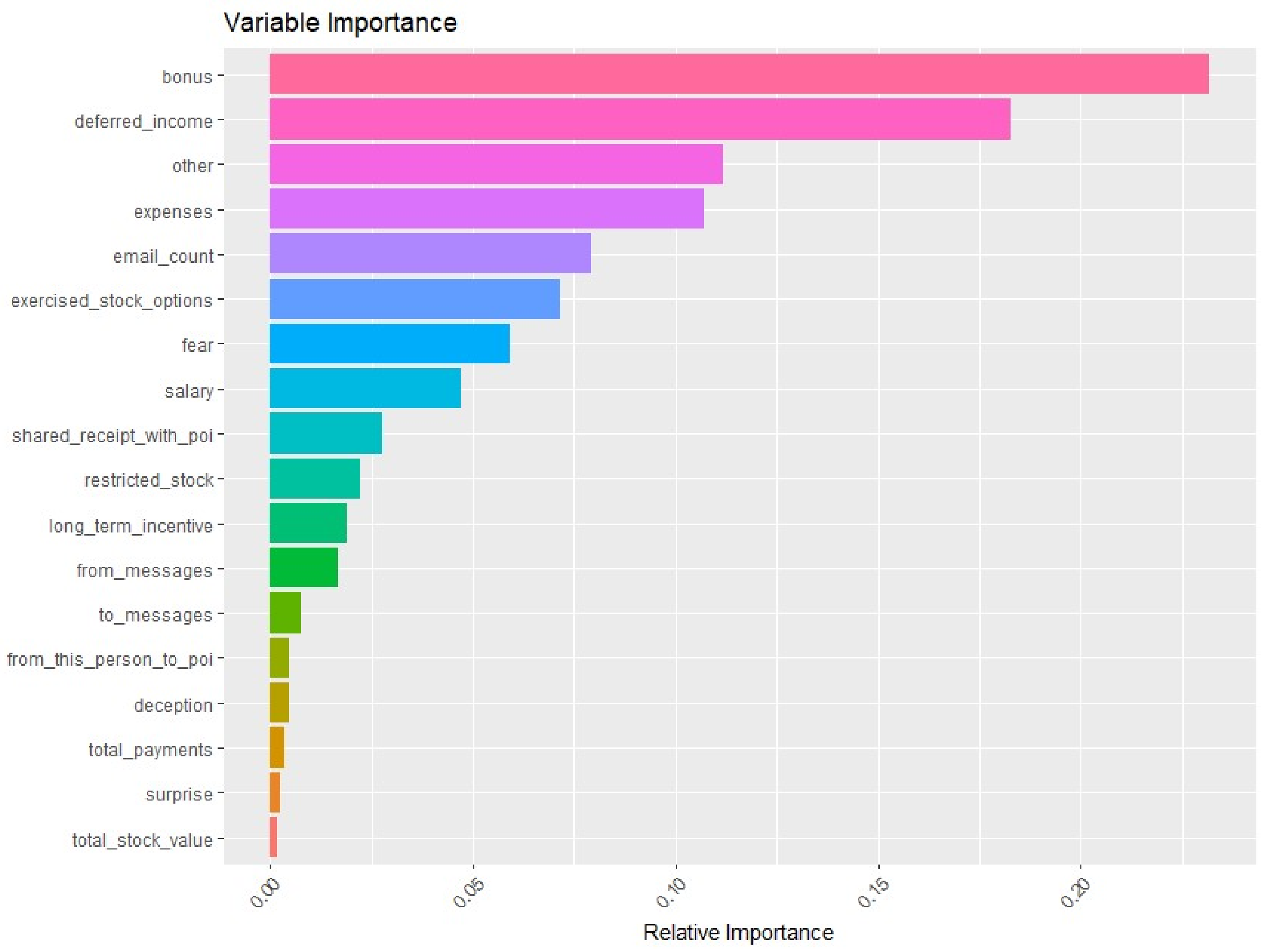

**Figure 12 Variable Importance for Predicting Persons of Interest from Financials and Emails**

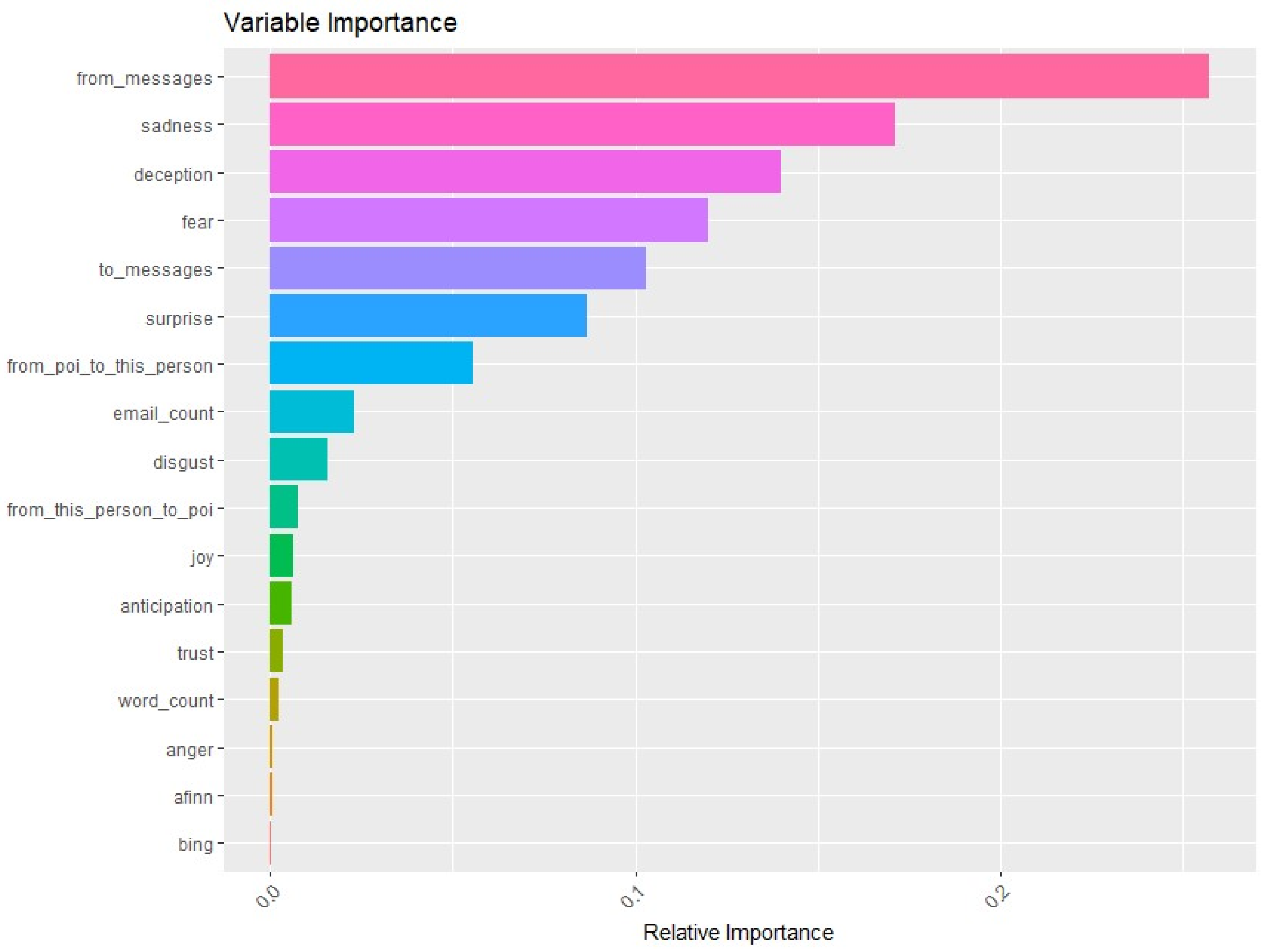

**Figure 13 Variable Importance for Predicting Persons of Interest from Emails**

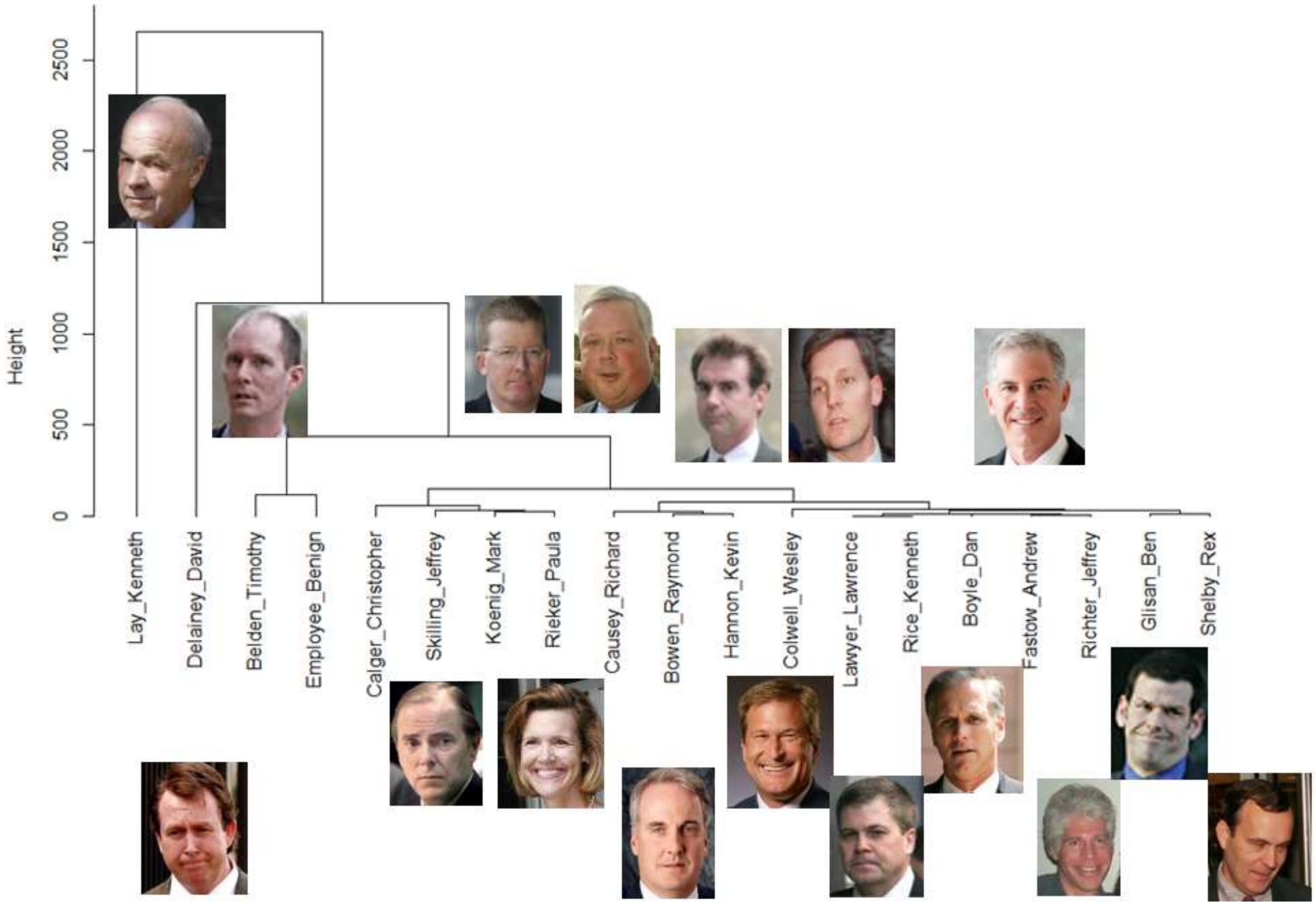

**Figure 14 Email Clusters by Employee and Stylistic Similarities.** *The emails from Ken Lay stand out as unique among other persons of interest. Tim Belden writes the most like a generic benign employee.*

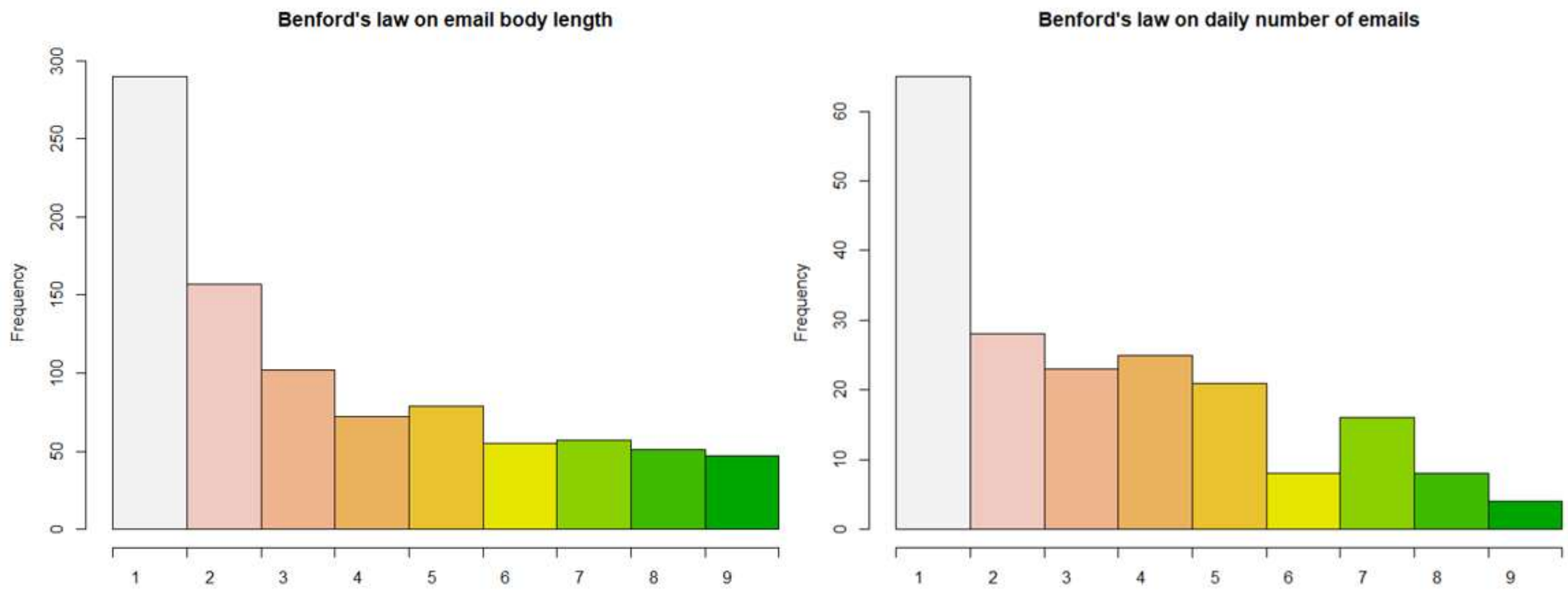

**Figure 15. Validity Check for Benford's Law Behavior in Subset of Enron Email Corpus.** *Such power law behavior reveals that Enron email follows expected statistical over-representation of small numbers.*